\documentclass[aip,graphicx,prb,longbibliography,reprint]{revtex4-1}

\usepackage{graphicx}

\draft 

\begin{document}

\title{Ballistic Deposition of Nanoclusters} 



\author{Jeffrey G. Ulbrandt}
\author{Yang Li}
\author{Randall L. Headrick}
\email[]{rheadrick@uvm.edu}
\affiliation{Department of Physics and Materials Science Program, University of Vermont, Burlington VT 05405}


\date{\today}

\begin{abstract}
Nanoporous thin-films are an important class of materials, offering a way to observe fundamental surface and bulk processes with particles larger than individual atoms, but small enough to interact significantly with each other through mechanisms such as stress and surface mobility. In-Situ X-ray Reflectivity and Grazing Incidence Small Angle X-Ray Scattering (GISAXS) were used to monitor thin-films grown from Tungsten Disilicide (WSi$_2$) and Copper (Cu) nanoclusters. The nanoclusters ranged in size from 2 nm to 6 nm diameter and were made by high-pressure magnetron sputtering via plasma-gas condensation. X-Ray Reflectivity (XRR) measurements of the film at various stages of growth reveal that the resulting films exhibit very low density, approaching 15\% of bulk density. This is consistent with a simple off-lattice ballistic deposition model where particles stick at the point of first contact without further restructuring. Furthermore, there is little merging or sintering of the clusters in these films.

\end{abstract}


\maketitle 



\begin{figure*}
\includegraphics[width=6.0 in]{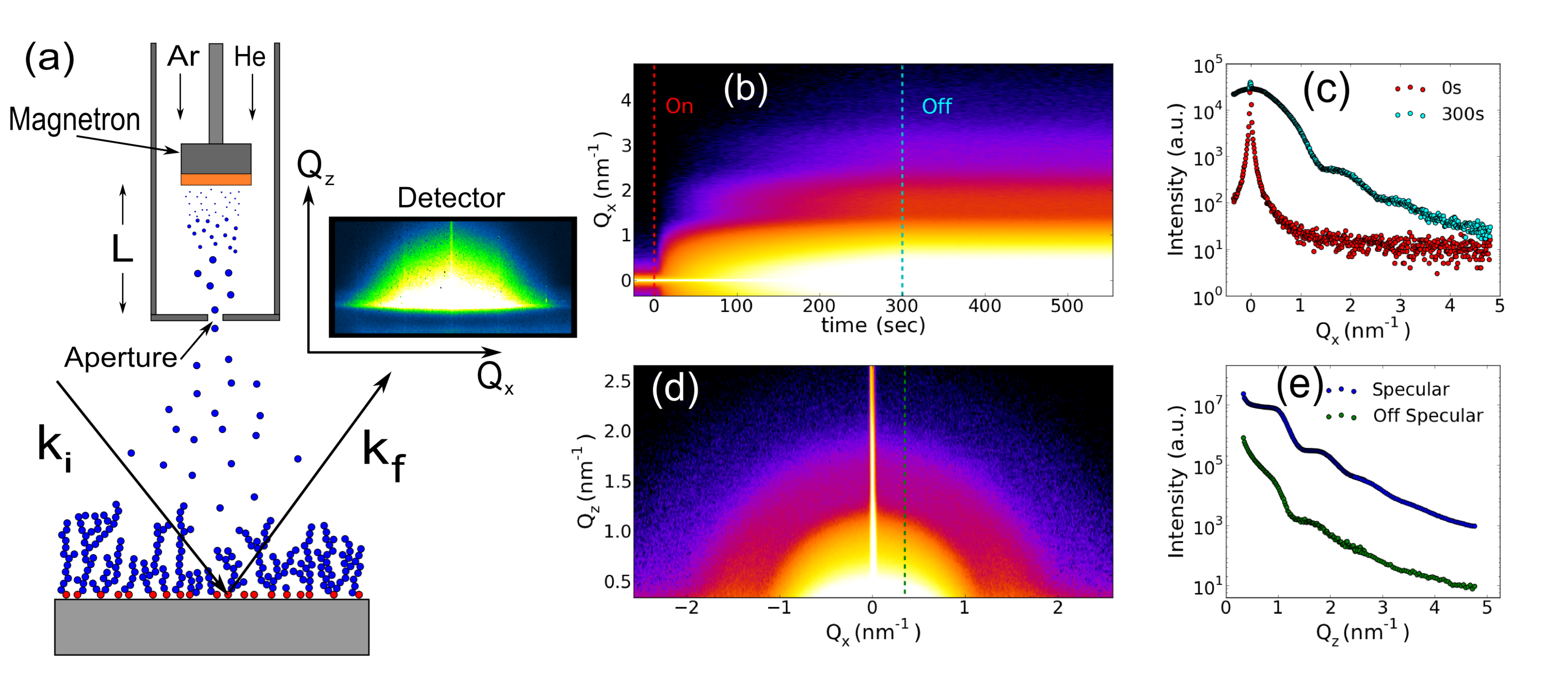}
\caption{\label{Experiment} (a) Schematic of the Nanocluster Source during deposition experiment. The cluster are generated by high-pressure magnetron sputtering. The clusters are guided through a small aperture by the gas flow into the deposition chamber where they are deposited on the substrate. Real-Time X-Ray Scattering data is collected during the deposition process. Two scattering geometries were employed to capture in-plane (Q$_x$) and out of plane (Q$_z$) information throughout the deposition. (b) Analyzed Data from the GISAXS geometry showing the time evolution of scattering along the Q$_x$ direction. The dashed lines mark the times when the deposition was started and ended. The development of scattering due to the spherical clusters can be seen. (c) The GISAXS data before and after the deposition. (d) Analyzed Data from the Reflectivity Geometry. (e) Crossection through the image in (d) showing the specular scattering and diffuse scattering as a function of Q$_z$. The green data corresponds to the green dashed line in (d).}
\end{figure*}

\begin{figure*}
\includegraphics[width=6.00 in]{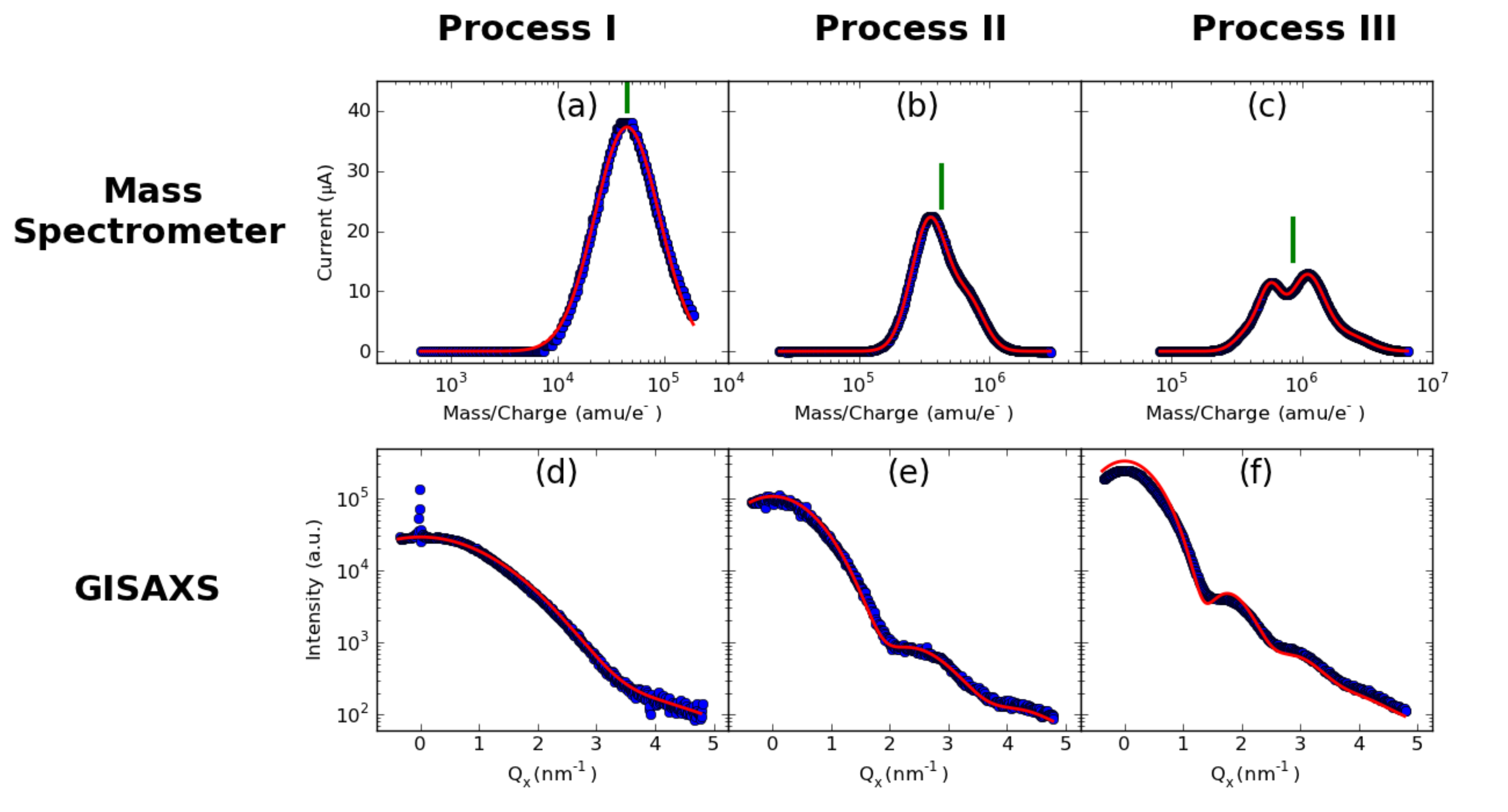}
\caption{\label{Mass_GISAXS} (a)-(c) Mass Spectrometer curves for three different size processes. The red lines are fits using a log-normal distribution. The green marks are the peak mass which, for when multiple peak were present,was calculated as a weighted mean of the individual peak masses. (d)-(f) GISAXS curves for the same processes along with fits using a log-normal distribution of diameters. The peak mass and peak diameter values were then used to calculate the density of the clusters for the three processes.}
\end{figure*}

\begin{table*}
\caption{\label{table1} Nanocluster process parameters, mass, diameter, and density for L = 150 mm.}
\begin{ruledtabular}
\begin{tabular}{cccccccccc}
Process & \multicolumn{2}{c}{Argon Flow \& Pressure} & \multicolumn{2}{c}{Helium Flow \& Pressure} & Power & Mass & Diameter & \multicolumn{2}{c}{Density}\\
& (sccm) & (Torr) & (sccm) & (Torr) & (W) & (amu) & (nm) & (amu/nm$^3$) & (g/cm$^3$)\\
\hline
WSi$_2$ - I & 20 & 0.243 & 1 & 0.394 & 25 &  5.02$\times$10$^4$ & 2.25 & 8.41$\times$10$^3$ & 13.95\\
WSi$_2$ - II & 50 & 0.550 & 0 & 0 & 50 &  3.51$\times$10$^5$ & 4.0 & 1.04$\times$10$^4$ & 16.72\\
WSi$_2$ - III & 50& 0.550 & 0 & 0 & 25 &  8.61$\times$10$^5$ & 5.6 & 1.45$\times$10$^4$ & 15.67\\
Copper & 100 & 1.0 & 0 & 0 & 100 &  3.14$\times$10$^5$ & 4.9 & 5.02$\times$10$^3$ & 8.47\\
\end{tabular}
\end{ruledtabular}
\end{table*}

Nanoporous Thin-Films have been of great interest in recent years due to their unique properties, many of which originate from the large surface area to volume ratio. One important application has been the use of nanoporous Titanium Dioxide (TiO$_2$) in dye-sensitized solar cells.\cite{Oregan1991} The cluster size and porosity of the films have been shown to be key parameters in the overall efficiency of the cells.\cite{Barbe1997} The deposition process has a large impact on the fianl structure. Other applications of nanostructures, using Tungsten Silicides, show promise for use in super capacitors.\cite{Banis2013}. Another application is in thermoelectric materials.\cite{Petermann2015,Mingo2009}  In this paper, we show that growing films by Ballistic Deposition of Nanoclusters result is nanoporous films with large surface area, which may be well suited for such devices.

The nanocluster source used in this study is based on Plasma Gas Condensation (PGC) in which clusters are nucleated and grown by thermalization of a high temperature supersaturated vapor. A source of this type was first developed by Haberland \cite{haberland_1991_newsource} and has been used by numerous groups since.\cite{Hihara1998,Drabik2011,Meinander2009,UrbanIii2002} The vapor is generated by magnetron sputtering. Much of the existing literature has dealt with single element materials, however this process is well equipped to make multi-element compounds or alloys.\cite{martinez_2012} We have used the source to grow thin-films from copper clusters as well as tungsten disilicide (WSi$_2$) clusters. Films grown using such a nanoclusters source result in very low density films which follows a very simple Ballistic Deposition model.

Ballistic Deposition without restructuring is one of the simplest deposition models. In this model, particles are deposited at normal incidence to the substrate, and when a particle comes in contact with the surface or another particle it stops and sticks. The resulting film structure is a very loose tree-like structure with few contact points between particles. Numerical simulations of this process have been carried out for mono-dispersed spherical particles which show that the packing fraction of the spheres in the bulk of the film, the volume occupied by the spheres divided by the total volume of space, is 0.1465.\cite{Jullien1987} In addition, there is very few points of contact between particles, the average being two, and the first several layers of particles have a higher density than the bulk of the film.\cite{Lubachevsky1993, Meakin1986, Lubachevsky1996}


The nanocluster source was custom built along the design set by Haberland\cite{haberland_1991_newsource} and is shown in Fig. \ref{Experiment} (a). The magnetron source was mounted inside a water-cooled stainless-steel tube, called the drift tube, with a 3 mm aperture at one end. The magnetron has a linear translation mechanism allowing the target to aperture distance to be adjusted from 0 mm to 200 mm. This was then installed in a UHV chamber pumped by a high-volume turbomolecular pump. The aperture served to allow the inner chamber to reach high enough pressure to allow the gas condensation process to occur. Additionally, at high enough pressures to enter the fluid flow regime, it collected the clusters into a beam which could then be directed into either a mass spectrometer or onto a substrate. The key parameters affecting the size and size distribution of the clusters were Target to Aperture distance (L), Gas Pressures, and Magnetron Power. The pressure ranged from 50 mTorr to 1 Torr and L was fixed at 150 mm for all the processes. Two gases, Argon and Helium, were used depending on what size of cluster was desired. For larger clusters, only Argon gas was used. To get a smaller diameter cluster, a lower Argon pressure was used and Helium was added to raise the overall pressure into the fluid regime.

The materials used in this study were WSi$_2$ and Copper. Three processes, with different sizes from 2 nm to 6 nm, were developed for WSi$_2$ and were labeled process I, process II, and process III from smallest to largest. Process parameters are shown in Table \ref{table1}.

In order to measure the mass distribution of the cluster processes, a MesoQ mass spectrometer was purchased from Mantis Deposition Ltd. The mass spectrometer was able to scan from 350 amu to 10$^6$ amu. A property of clusters generated by PGC is that the majority of the clusters have an inherent negative charge. This allowed the use of the mass spectrometer without an ionization stage. In order to improve the vacuum inside the mass spectrometer, the main chamber and spectrometer chamber were isolated by another aperture of 8 mm diameter. The spectrometer chamber was pumped by a 300 l/s turbomolecular pump.

For In-situ X-ray scattering experiments, the drift tube was installed on a custom UHV chamber at beamline X21 of the National Synchrotron Light Source (NSLS) at Brookhaven National Lab. The Scattering was performed at an energy of 10 keV with an X-Ray spot size of 0.5 mm by 1 mm. The detector was a Pilatus 100k pixel array area detector. In order to measure both in-plane and out of plane film structures, two scattering geometries were used, Grazing Incidence Small Angle X-Ray Scattering (GISAXS) and X-Ray Reflectivity (XRR). For Real-Time measurements of the deposition process, GISAXS was performed by setting the exit angle at the critical angle of the substrate. The incidence angle was set above the critical angle to keep the specular reflection off the detector. At various mean film thicknesses the deposition was stopped and XRR was performed. In this geometry the incident and exit angles were set equal and swept from 0.1 to 5 degrees. In addition to the specular beam, diffuse scattering along the in-plane direction (Q$_x$) was measured at each incident angle. This allowed for a full mapping of the diffuse scattering in Q$_x$ and Q$_z$ at discrete thicknesses. 


The density of the clusters was determined from the cluster mass and cluster diameter. The mass spectrometer was used to characterize the mass distribution for all processes. For clusters grown in a vapor absorption process, the cluster size distribution follows a lognormal distribution.\cite{Soederlund1998} A simple form of a log-normal distribution is given by:

\begin{equation}
\Phi(x, \mu, \sigma) = \frac{1}{\sigma\sqrt{2 \pi}} 
\exp\left[
-\frac{(\ln x - \ln \mu)^2}{2\sigma^2}
\right], x>0
\end{equation}

In this form of the equation the parameter $\mu$ represents the peak value (mode) of the distribution. Since the mass of the clusters goes as the cube root of the radius, it follows that the mass distribution is also lognormal. Mass Spectrometer curves for the three processes are shown in Fig. \ref{Mass_GISAXS} (a)-(c). The plot is semi-log x and shows that the distribution appears normal, as is the case for log-normal distributions. The red lines are fits using Eq. 1. It was found that the larger size processes had multiple peaks develop indicating a more complex nucleation and growth process. A more detailed view of the fits using multiple distributions is shown in the supplemental materials. The peak mass for each process is shown in Table \ref{table1}. The mean value of the peak mass for Process III was used since it showed two prominent peaks.

The cluster diameter was determined from the GISAXS measurement during early stages of growth, where the coverage of the surface was low. At this stage there should be little contact between clusters and the clusters will be randomly distributed on the surface. In general, the scattering intensity will be given by I = S($\vec{q})|F(\vec{q},r)|^2$, where $F(\vec{q},r)$ is the form factor for a spherical particle of radius r, and $S(\vec{q})$ is the interference function relating to particle positions. As the deposition process is random, and surface diffusion is assumed to be low, there is no long range order in the arrangement of clusters and the GISAXS intensity is proportional to the particle size only, i.e. ignore any contribution from $S(\vec{q})$. Finally, since there is a distribution of particle sizes, the overall scattering intensity is given by I = $\sum\Phi(r)|F(\vec{q},r)|^2 /\sum\Phi(r)$, where $\Phi(r)$ is the distribution of particle sizes. The  data is shown in Fig. \ref{Mass_GISAXS} (d)-(f), with fits in red, using a lognormal distribution for the particle sizes. The peak diameter for each process was taken from the parameter $\mu$ and is shown in Table \ref{table1}.

The density of the particles for each process is summarized in Table \ref{table1}. It is interesting to note that for all the WSi$_2$ processes, the density was quite a bit higher than the bulk density of WSi$_2$ (9.3 g/cm$^3$). The density is much closer to another stable phase of tungsten silicide, W$_5$Si$_3$ (14.55 g/cm$^3$). As a control, the copper density was calculated and is close to the bulk value (8.9 g/cm$^3$).


Next, the structure of the films was analyzed. X-Ray reflectivity at various stages of deposition revealed how the structure of the film changed with mean thickness. Fig. \ref{Experiment} (d) shows an image constructed from the specular and diffuse scattering during a late stage deposition reflectivity scan. The image is interesting in that it shows both the diffuse scattering from the spherical clusters, which is spread out in reciprocal space, and the specular rod which is confined to Q$_x$ = 0. More interesting is comparing the specular scattering to the off-specular scattering as shown in Fig. \ref{Experiment} (e). The off specular diffuse scattering shows fringes due the the spherical clusters that make up the film. Looking at the specular scattering we also see the same fringes. Clearly, as the specular intensity is several orders of magnitude greater than the diffuse intensity, these fringes are part of the specular scattering.

\begin{figure}
\includegraphics[width=3.50 in]{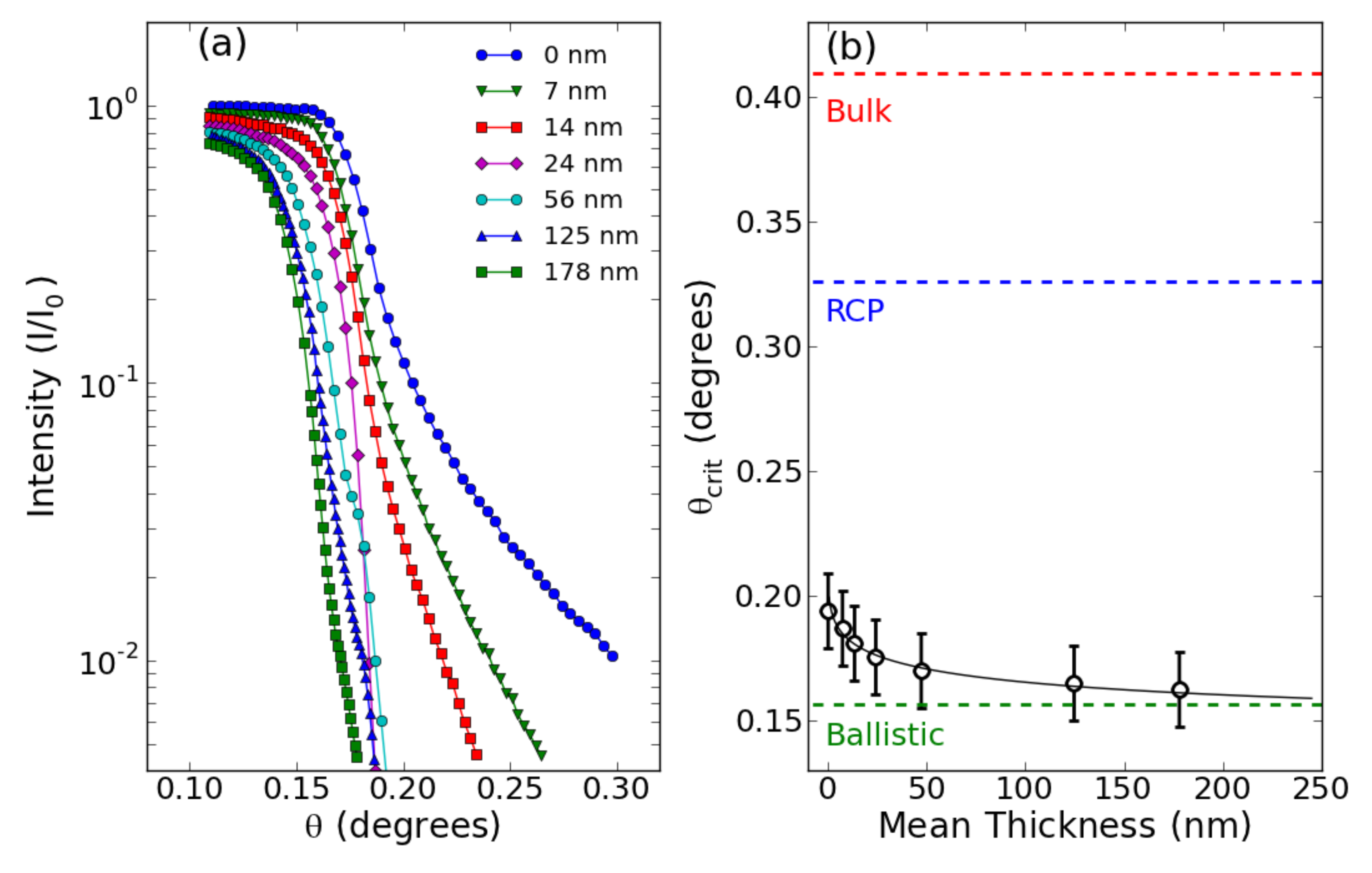}
\caption{\label{Critical} Evolution of the Critical Angle with thickness. The raw data at several mean thickness is shown in (a). The critical angle as a function of mean thickness is shown in (b). The dashed lines are the expected critical angle for different packings of clusters. The data is clearly converging towards the ballistic model.}
\end{figure}

The specular reflectivity contains information about the average electron density in the direction normal to the surface. The critical angle of the curves is related to the average electron density by: $\theta_c = \sqrt{4 \pi \rho r_0}/k$.\cite{Als-Nielsen2011} Fig. \ref{Critical} (a) shows the evolution of the critical angle with mean thickness. For a perfectly smooth curve the intensity drops abruptly at the critical angle. However for a rough interface, the curve is rounded out near the critical angle. The value of the critical angle was taken as the location of the steepest slope on the curve. Fig. \ref{Critical} (b) shows the critical angle versus mean thickness. It is clear that the critical angle is falling to lower Q indicating an even lower density than the substrate. Several expected critical angles for different packing fractions are shown as dashed lines. These represent a continuous film, random close pack (RCP), and Ballistic deposition without restructuring. The data is converging on the prediction of the ballistic deposition model.

\begin{figure}
\includegraphics[width=3.50 in]{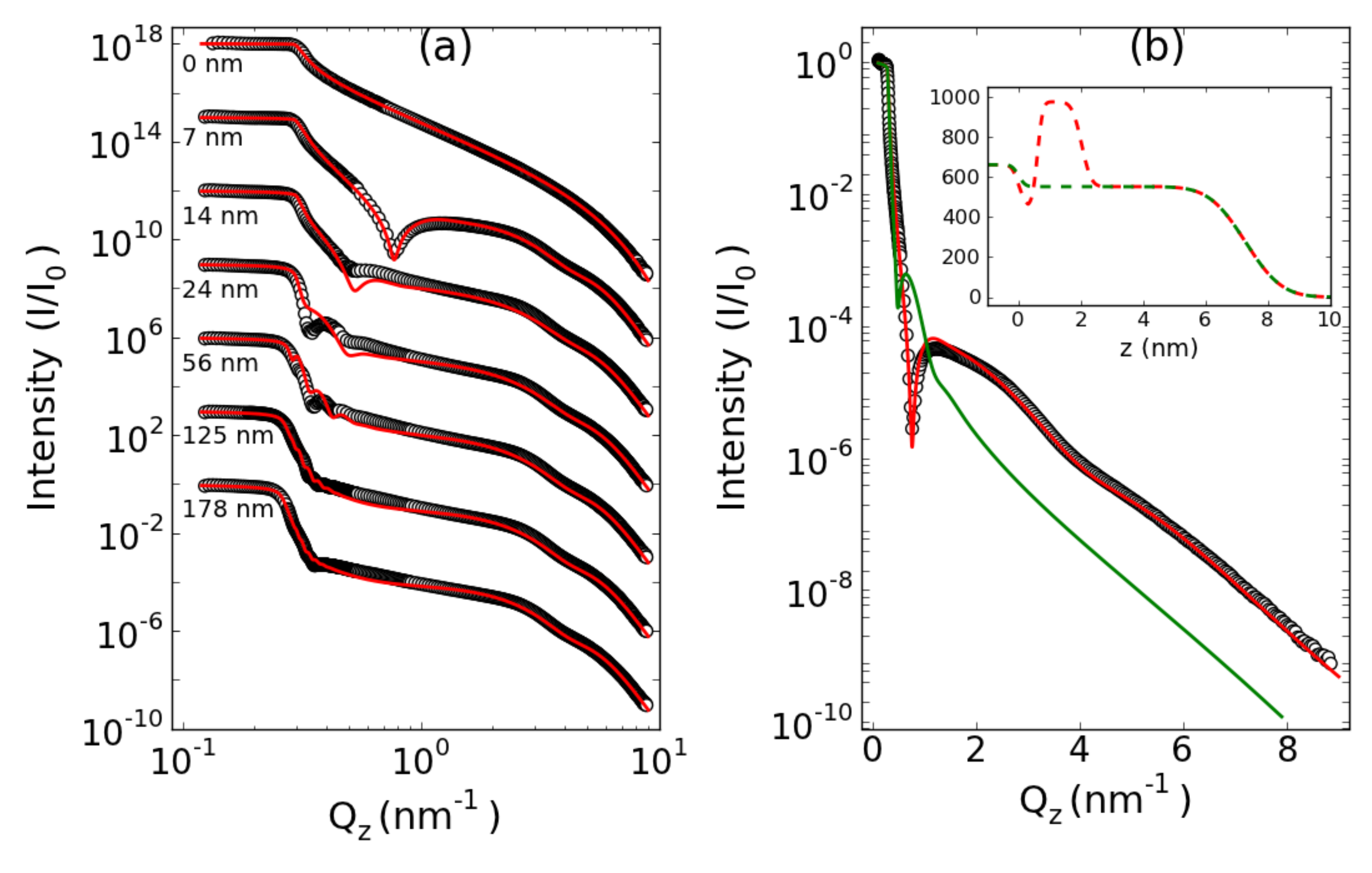}
\caption{\label{Reflectivity} (a) Reflectivity scans for different mean thicknesses along with fits for Process I. Scattering at high Q develops early in the deposition due to the layer of particles in contact with the surface. This scattering remains fixed through the remainder of the deposition. Kiessig fringes develop at low Q, though are heavily damped due to the roughness of the top layer. (b) The effect of the first layer of clusters on the reflectivity. The data is the 7 nm mean thickness from Process I. The green line represents a single layer film with a mean thickness of 7 nm. The red line is a three-layer film of the same mean thickness, however the first two layer represent the layer of clusters in contact with the substrate. This model is a much closer fit of the data. The corresponding electron density for the two models is shown in the inset. This electron density is similar to the electron density predicted by ballistic deposition.}
\end{figure}

The specular reflectivity contains more information than just the average electron density of the film. As seen in Fig. \ref{Reflectivity} a. there are several features in the curves which are related to the structure of the films. To evaluate these features, the curves were fit using a simple multilayer model and the Parratt formalism. The Parratt method is a dynamical theory based on scattering off continuous layers and also accounts for multiple scattering events.\cite{Parratt1954} This method is justified since the specular reflectivity is not sensitive to the in-plane structure of the film. Therefore, clusters on a surface appear as a continuous film, only with a lower density than the cluster density, which depends on the coverage of clusters. We also include a roughness factor for each layer which assumes a Gaussian roughness.

The evolution of the specular reflectivity has several interesting features which are unique to this deposition model. Typically when a thin film grows we see the appearance of Kiessig fringes, due to interference between scattered x-rays from the film surface and substrate. The spacing of the fringes is inversely proportional to the thickness of the film, which decrease as the film gets thicker. In our data we do see oscillations appear early in the deposition process. However, they stay fixed in q throughout the whole deposition. Only for the thicker layers do we see Kiessig fringes develop. The absence of Kiessig fringes during the early deposition is due to the roughness of the layer. The roughness is on the order of the thickness for the early stages of deposition which completely damps out the Kiessig fringes. At later stages, the thickness becomes much larger than the roughness, and Kiessig fringes appear, though heavily damped due to the roughness.

The features at high-q which remain static throughout the deposition are due to the layer of clusters which are in contact with the surface. In the ballistic deposition process, most of the clusters are at different heights which are uncorrelated and hence will average out. However, clusters in contact with the substrate are at the same height and hence to not average out. This layer will have a higher electron density than the rest of the layer. Fig. \ref{Reflectivity} b. shows the reflectivity for a single layer versus the reflectivity for a multilayer. The addition of this first layer fits the data quite nicely. The inset shows the corresponding electron density for each case.


We have shown that a simple ballistic deposition model fits the structure of nanocluster deposited thin-films. The clusters have few points of contact leaving most of the surface area exposed. An interesting question is to what degree the cluster merge on contact. For W$_2$, there appears to be very little merging and the clusters retain their spherical shape, as seen in the diffuse scattering (Fig. \ref{Experiment} (d)). In the copper cluster data, we do see evidence that the clusters are merging, though on a very slow time scale. This is shown in the supplemental material. Even with a significant degree of sintering, since the number of contact points is very low and density of clusters is very low, it is unlikely for the material to merge into a continuous thin film. A more detailed study of this sintering behavior is warranted for future studies.

%
%



%


\bibliography{Nanocluster_films_V3}

\end{document}